\documentclass[twocolumn,showpacs,amsmath,amssymb]{revtex4}
\usepackage{graphicx}
\usepackage{color}

\begin{document}

\title{Bi-axial shear of confined colloidal suspensions: the structure and rheology of the vorticity-aligned string phase}
\author{Neil Y. Lin$^1$, Xiang Cheng$^{1,2}$, Itai Cohen$^1$}
\affiliation{$^1$Department of Physics, Cornell University, Ithaca, New York 14853, $^2$Department of Chemical Engineering and Materials Science, University of Minnesota, Minneapolis, Minnesota 55455}
\date{\today}

\begin{abstract}
Using a novel bi-axial confocal rheoscope, we investigate the structure and rheology of sheared colloidal suspensions under confinement. Consistent with previous work  [X. Cheng \textit{et al., Proc. Natl. Acad. Sci. U. S. A.}, 2011, \textbf{109}, 63], we observe a vorticity-aligned string phase in moderate concentrated colloidal suspensions under uniaxial shear. Using bi-axial shear protocols, we directly manipulate the orientation and morphology of the string structures. Simultaneously, we measure the suspension rheology along both the flow and vorticity directions with a bi-axial force measurement device. Our results demonstrate that despite the highly anisotropic microstructure, the suspension viscosity remains isotropic and constant over the shear rates explored. These results suggest that hydrodynamic contributions dominate the suspension response. In addition they highlight the capabilities of bi-axial confocal rheoscopes for elucidating the relationship between microstructure and rheology in complex fluids.
\end{abstract}

\maketitle

\section{Introduction}
Strong shear flow can break the symmetry of isotropic phases in complex fluids and induce a highly anisotropic structure \cite{Solomon, Larson, Vermant2010}. Examples include the out-of-equilibrium isotropic-nematic phase transition in sheared liquid crystals \cite{Jean, Denis, Elmar}, the log-rolling cylindrical phase of micelles in amphiphilic films \cite{Arya, Beth}, and the shear-induced crystallization and buckling phases in concentrated colloidal suspensions \cite{Cohen, Haw}. The flow behavior of these fascinating phases has been intensively studied in rheological measurements using uniaxial shear \cite{Solomon, Larson, Vermant2010, Jean, Denis, Elmar, Carlos, Arya, Beth, Cohen, Haw}. However, this conventional method only measures the stress response  along the flow direction. It cannot reveal the directionally dependent rheological properties of anisotropic phases. For measuring these quantities, a bi-axial shear protocol with the superimposed perturbation method is required \cite{Farage,Vermant, orthogonal}. Combining such bi-axial shear protocols with direct imaging of the material microstructure holds the promise of enabling correlation between structure and anisotropic rheology in these structured materials. 

With that end in mind, we construct a bi-axial confocal rheoscope by combining a dual-directional shear cell with a bi-axial force measurement device. This apparatus allows for manipulating the formation and the orientation of shear-induced structures while simultanesously measuring their anisotropic mechanical response, and imaging their structure \cite{Besseling, Besseling2, xiang, Lehigh, Schmoller}. 

Here, we use this instrument to systematically investigate the structure and rheology of the vorticity-aligned string phase in sheared colloidal suspension under confinement. This phase is particularly interesting because it has a unique one-dimensional translational symmetry \cite{vermant2010a, vermant2010b, ackerson, ackerson1990, Laun, string, Osuji}, has been shown to arise from interesting hydrodynamic interactions between particles \cite{string}, and has potential applications to nano-fabrication and bio-analysis \cite{spin, Vermant2010, Baudry2006}. Until recently, prior research has focused on the string structure in bulk colloidal suspensions with visco-elastic polymeric solvents \cite{Solomon, Vermant2010, vermant2010a, vermant2010b, ackerson, ackerson1990, Laun}. However, it has been found that even in a simple Newtonian fluid, suspended colloidal particles can assemble into a string structure when confined to a narrow gap that is a few particle-diameters thick \cite{string}. In contrast to the flow-aligned string phase found in bulk colloidal suspensions with visco-elastic solvents, the confined strings orient along the vorticity direction normal to the plane of shear. The intriguing structure results from the unique interactions that arise from geometric constraint and hydrodynamic particle-particle and particle-wall couplings all of which are enhanced by confinement \cite{string}. Although the structure of the vorticity-aligned string phase under confinement has been reported \cite{string}, the flow behavior and stress response of this anisotropic phase are still poorly explored. With our bi-axial confocal rheoscope, we directly measure the rheology of the string phase along different directions. Furthermore, we demonstrate direct control over the isotropic-to-string phase transition using bi-axial shear flows.

\begin{figure}
\includegraphics[width=0.45\textwidth]{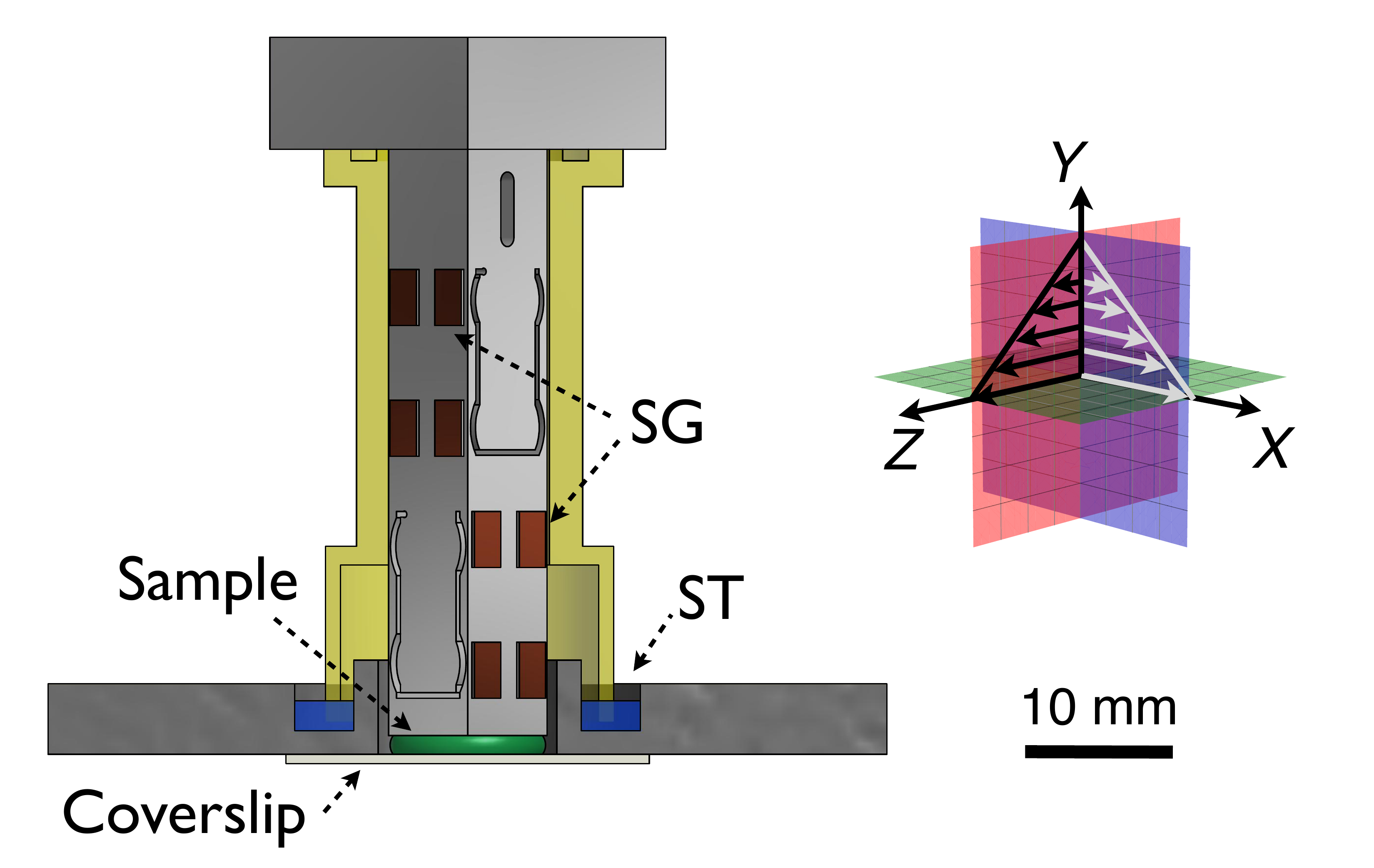}
\caption{Three dimensional schematic of the bi-axial force measurement device (left) and coordinate definition (right). The abbreviations SG and ST stand for strain gauges and the solvent trap, respectively. The lower strain gauges measure the stress response along the $X$-axis and the upper ones measure the response along the $Z$-axis. The gap between the top plate (silicone wafer) and the bottom plate (coverslip) is exaggerated for clarity. The gap separation in the experiments is 9.0$\mu$m. The shear flow is separately imposed along the $X$ and $Z$ axes by moving the bottom plate using a multi-axis piezoelectric actuator.}
\label{fig:fig_1}
\end{figure}

\section{Experimental method}
The schematic of the bi-axial shear cell is illustrated in Fig.~\ref{fig:fig_1}. For each experimental run, 10$\mu l$ of suspension are loaded in a gap consisting of a coverslip and a silicon wafer with a 9$\mu$m separation between them. Both plates are adjusted to be parallel within 0.0075$^\circ$ by turning three set screws. The coverslip is coupled to a multi-axis piezo (PI P-733) that can generate movements along all directions to apply bi-axial shear flows. The silicon wafer is attached to a bi-axial force measurement device (FMD) so that the shear stresses $\sigma_{xy}$ and $\sigma_{zy}$ are measured simultaneously. Here, $X$, $Y$, and $Z$ correspond to the flow, gradient, and vorticity axes of the primary shear flow. In the FMD, eight foil gauges - four for each direction - are wired as two independent Wheatstone bridges that enable stress measurement. All signals measured by the FMD are amplified by signal conditioning amplifiers (Vishay 2310B) then digitized for Fourier analysis. By mounting this bi-axial shear cell on a fast confocal microscope (Zeiss LSM 5 Live), we also image the microstructure while the suspension is sheared and the stress response is measured. 

Our sample is comprised of silica particles with diameter $a=1.3\mu$m suspended in a 1:4 water-glycerin mixture. The solvent has a viscosity $\eta=0.06$Pa$\cdot$s and a refractive index of 1.442 that matches that of the particles. The suspension volume fraction is 0.39. We add 1.25mg/ml of fluorescein sodium salt to dye the solvent for confocal imaging. 

The bi-axial shear flow imposed can be divided into two oscillatory shear flows 
\begin{eqnarray}
\vec{\gamma}(t)= \vec{\gamma}_1\sin(\omega_1 t ) + \vec{\gamma}_2 \sin(\omega_2 t + \delta).
\label{eq:2D}
\end{eqnarray}
Here, $\vec{\gamma}$ corresponds to the strain amplitude, $\omega$ corresponds to the shear frequency, and $\delta$ corresponds to the phase angle difference between the primary flow, indicated by the subscipt 1 and secondary flow, indicated by the subscript 2. 

\section{Results}

\subsection{Phase angle dependence: structure}

\begin{figure*}
\centering
\includegraphics[width=0.9\textwidth]{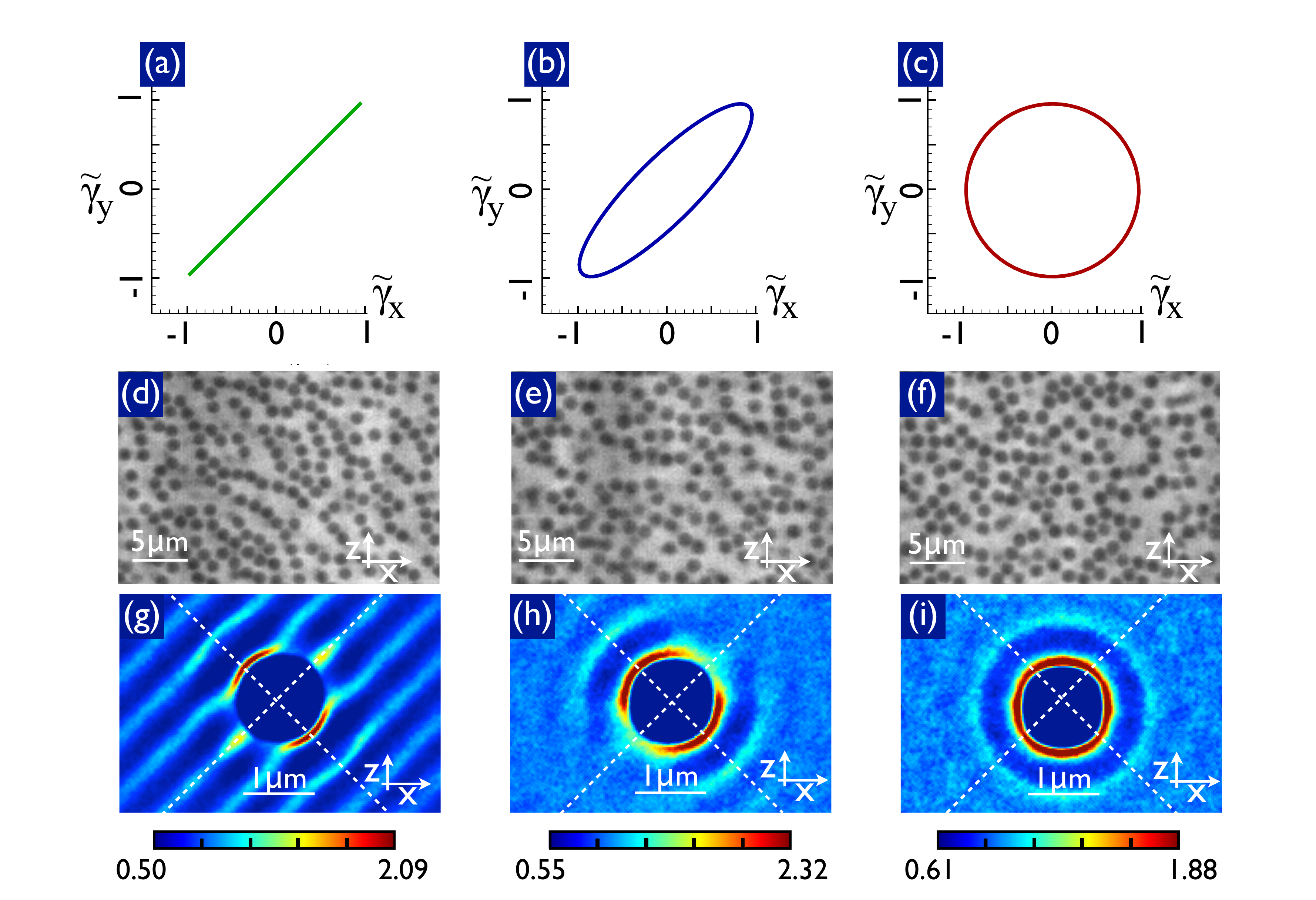}
\caption{Suspension microstructure for three phase angles $\delta=0, \pi/6$ and $\pi/2$.  Lissajous curves for the normalized shear strains $\tilde{\gamma}_y$ versus $\tilde{\gamma}_x$ are plotted in (a)-(c), where (a) (b) and (c) correspond to $\delta=0, \pi/6$ and $\pi/2$ respectively. The confocal images of the suspension with $\delta=0, \pi/6$ and $\pi/2$ are shown in (d), (e), and (f) respectively. The corresponding pair correlation functions $g(\vec r)$ of the particle distribution are shown in (g), (h) and (i). Each $g(\vec r)$ image represents a measurement averaged over 20 shear cycles. The dashed lines are guides for the eye and denote the $45^\circ$ and the $135^\circ$ orientations. The color ranges in the density plots are chosen to emphasize the structural features.}
\label{fig:fig_2}
\end{figure*}

In the phase angle experiment, we set two perpendicular flows at $|\vec{\gamma}_1|= |\vec{\gamma}_2|=2.50$, $\omega_1=\omega_2=31.4s^{-1}$ and vary the phase angle $\delta$ over the range $0\leq \delta \leq \pi$. Thus the shear rate along each axis is $\dot\gamma_{1,2}=78.5$s$^{-1}$, which corresponds to Pe$=4.73 \times 10^3$. Here, Pe=$6 \pi \eta_0 \dot{\gamma} a^3/(8k_B T)$ is the P$\acute{\rm{e}}$clet number that characterizes the ratio of the shear rate $\dot\gamma$ to the relaxation rate $1/\tau_s$ of the sample.

Fig.~\ref{fig:fig_2}(a), (b) and (c) show the Lissajous curves for the imposed flows where $\delta=0, \pi/6$ and $\pi/2$ respectively.  For $\delta=0$, the shear strain trajectory is linear and aligned at 45$^\circ$ to the $X$-axis. This linearly polarized shear flow is exactly the same as the uniaxial shear flow except with a different orientation (Fig.~\ref{fig:fig_2}(a)). In Fig.~\ref{fig:fig_2}(b) and (c), $\vec{\gamma}(t)$ is elliptically polarized with $\delta=\pi/6$ and circularly polarized with $\delta=\pi/2$ respectively.

Previous measurements have shown that the string structures are most pronounced near the boundaries \cite{string}. Thus, for each phase angle, we image the colloidal particles in the second layer 2.5$\mu$m below the top stationary plate. We find that as $\delta$ changes from 0 to $\pi$/2 the suspension structure transitions from a string morphology to one that is isotropic (Fig.~\ref{fig:fig_2}(d-f)). To illustrate this transition we calculate the pair correlation functions $g(\vec r)$ (Fig.~\ref{fig:fig_2}(g), (h) and (i)). Here $g(\vec r)$ is the normalized probability of finding a particle at vector $\vec r$ with respect to another particle in the $X$-$Z$ plane. In Fig.~\ref{fig:fig_2}(d), we find that when the suspension is subjected to a linearly polarized shear flow, $g(\vec r)$ demonstrates a highly anisotropic distribution at its first peak and exhibits stripes at larger $\vec r$. These stripes along with the anisotropic distribution of particle densities confirm that particles align along the vorticity direction and form string structures under uniaxial shear (Fig.~\ref{fig:fig_2}(g)). This finding is consistent with previous results \cite{string}. As $\delta$ increases to $\pi/6$, the first peak of $g(\vec r)$ maintains a similar shape but with a broader peak width. Most stripes disappear and the anisotropy of the second and the third peaks of $g(\vec r)$ significantly decreases (Fig.~\ref{fig:fig_2}(h)).  Finally, for $\delta=\pi/2$, $g(\vec r)$ is isotropic indicating that the suspension is characterized by liquid-like order Fig.~\ref{fig:fig_2}(i).

\begin{figure}
\includegraphics[width=0.45\textwidth]{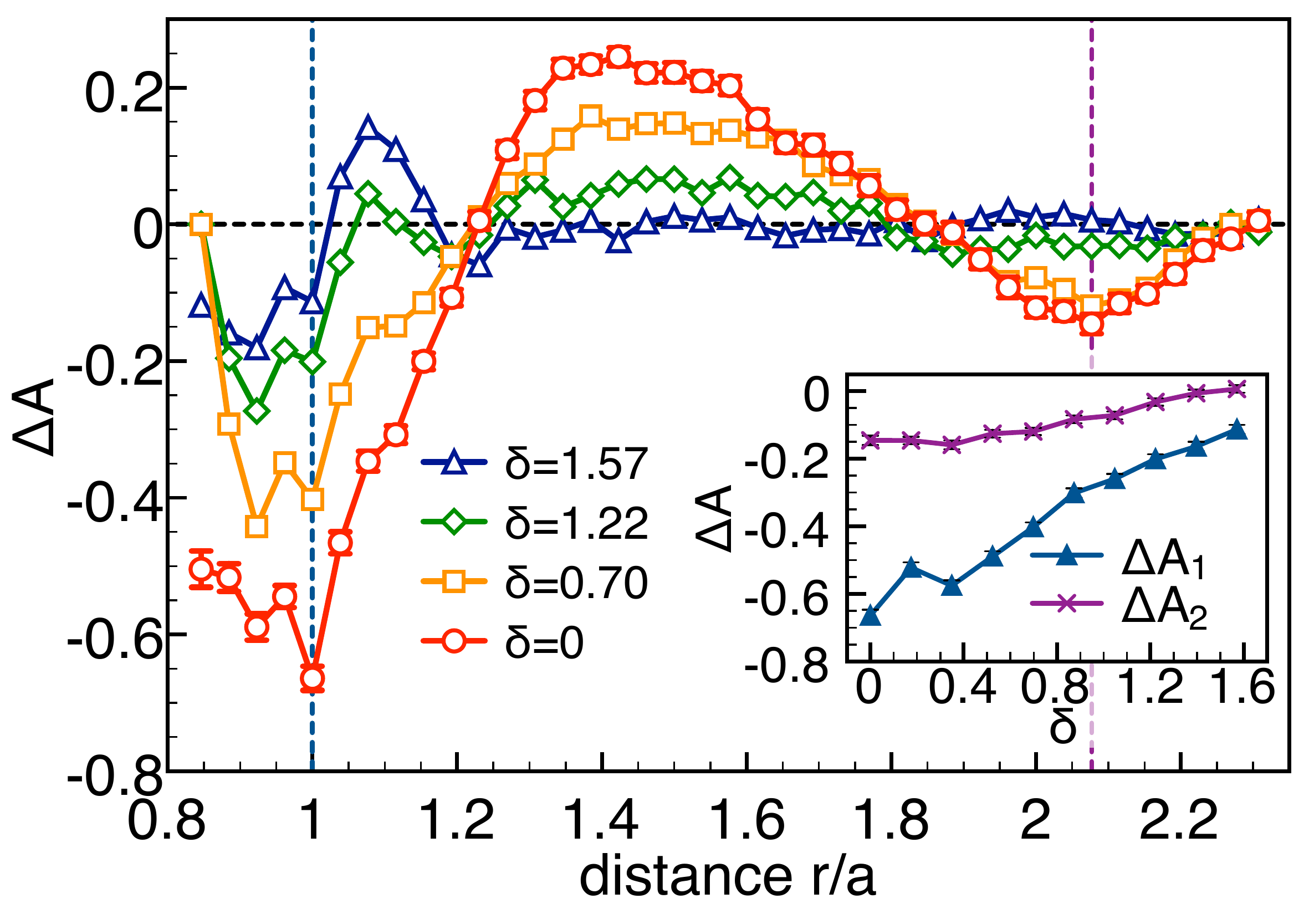}
\caption{Alignment factor $\Delta A$ at four different $\delta$. At small phase angles, $\Delta A$ oscillates and has two negative maxima at $r/a=1$ (vertical left dashed line) and $r/a=2.1$  (vertical right dashed line). The inset plots the alignment factors $\Delta A_1$ and $\Delta A_2$ versus $\delta$.}
\label{fig:fig_3}
\end{figure}

To determine the degree of order we circularly integrate $g(\vec{r})$ weighted by $\cos(2 \theta)$ and construct the order parameter $\Delta A$.  

\begin{eqnarray}
\Delta A(r) = \int_{0}^{2 \pi} g(r,\theta) \cos(2\theta) d\theta
\label{eq:order}
\end{eqnarray}
where $g(r,\theta)$ is the pair correlation function in polar coordinate, and $\theta$ is the angle between $\vec{r}$ and the flow direction. This order parameter is the real space analogue of the alignment factor, which has been used to quantify anisotropic suspension structures \cite{vermant2010a, Maranzano2002}. $\Delta A$ is positive for flow-aligned structures and negative for vorticity-aligned structures. We calculate $\Delta A$ versus $r/a$ for ten different values of $\delta$, and plot the data for four values in Fig.~\ref{fig:fig_3}. We find that for $\delta=0$,  $\Delta A$ has two negative extremes located at $r/a=1.0$ and $2.1$, corresponding to the vorticity-aligned structures. The sign of this functional form is opposite to the one found for flow-aligned structures \cite{vermant2010a, Maranzano2002}. As $\delta$ increases, the oscillation of $\Delta A$ decreases and becomes flat at $\delta=\pi/2$.

 We measure $\Delta A$ at the first and second negative peaks $\Delta A_1$ and $\Delta A_2$ and plot them versus $\delta$. We find that both $\Delta A_1$ and $\Delta A_2$ decay with increasing phase angle as illustrated in the inset of fig.~\ref{fig:fig_3}. This gradual decrease indicates that the transition from the string phase to isotropic phase is continuous. We also find that $\Delta A_2$ is smaller than $\Delta A_1$ for all $\delta$. The ratio $\Delta A_1 / \Delta A_2$ characterizes the probability ratio of finding a two-particle chain to that of a three-particle chain aligned along the vorticity direction. In combination with the real space images, these data demonstrate that as $\delta$ is decreased, the particles form chains that are both long and kinked.

These trends are valid so long as the shear period is much smaller than the relaxation time. Stokesian Dynamics simulations\cite{diffusion,diffusion2,Sierou} and previous experimental results\cite{string}, show that for the applied shear rate, the diffusivity is enhanced by a factor of 200. Thus, the time for a sheared sample to reach steady state is around $\tau_s/200\sim 0.3$s where $\tau_s = 60$s is the relaxation time for the quiescent suspension. This shear-induced relaxation enables particles to form strings within about one shear cycle. Thus, for the circularly polarized shear flow ($\delta=\pi/2$), if the radius of gyration and the period are both increased by a factor of 10 so that the shear rate is maintained, the string structures will still form but change their orientation with the imposed flow.

\subsection{Phase angle dependence: rheology}

\begin{figure}
\includegraphics[width=0.5\textwidth]{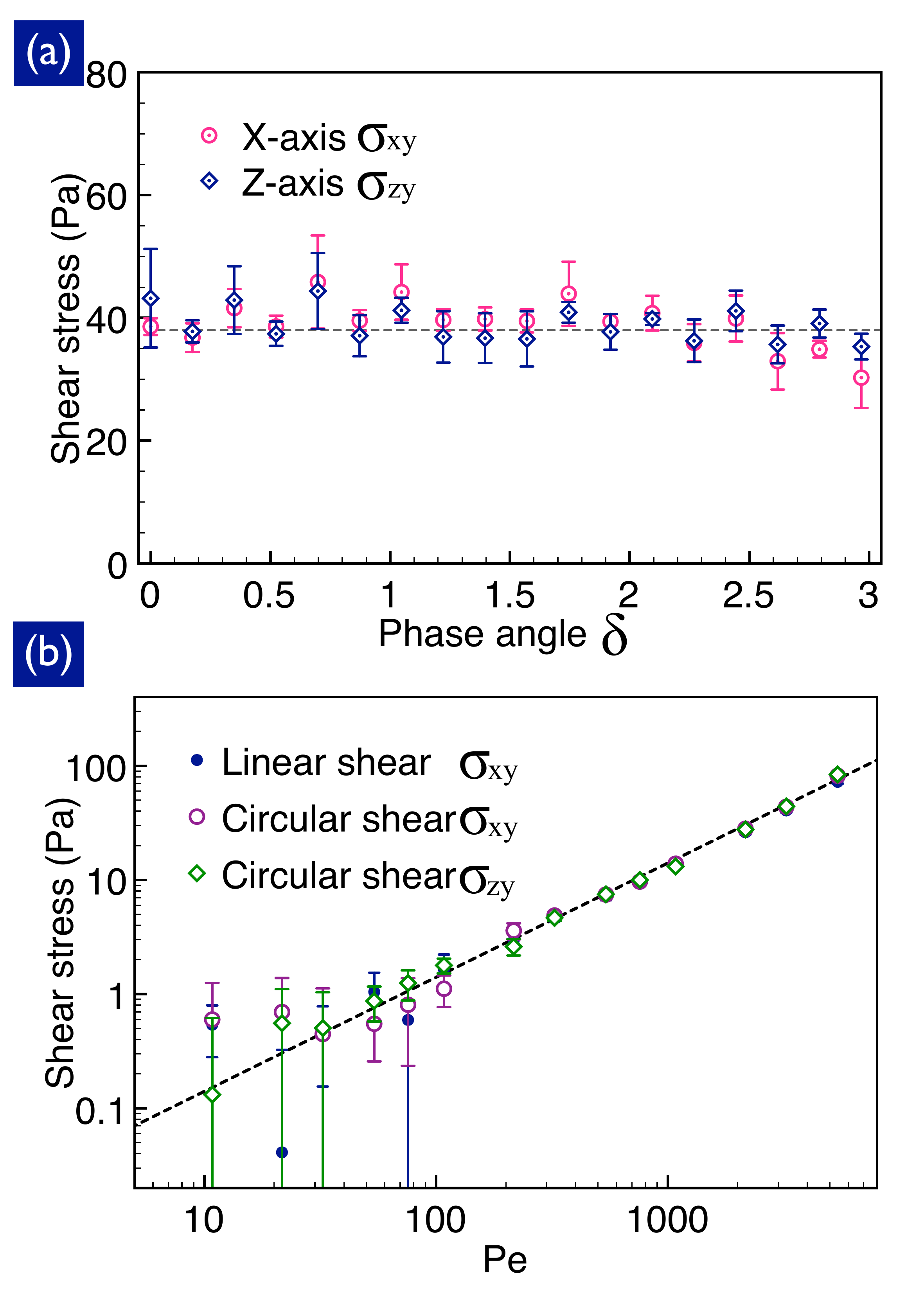}
\caption{Suspension stress response versus $\delta$ (a) and Pe (b). (a) The stress responses along the $X$-axis, $\sigma_{xy}$ and the $Z$-axis, $\sigma_{zy}$ are plotted versus $\delta$ with $\gamma_{1,2}=2.50$ and $\omega_{1,2}=31.4s^{-1}$. The dashed horizontal line indicates the mean value of the data. Each data point is averaged over five independent measurement runs with each run consisting of 500 cycles. (b) The stress response is plotted as a function of \rm{Pe} for linearly polarized and circularly polarized shear flows. The data are consistent with a Newtonian response as indicated by the linear fit (dashed line).}
\label{fig:fig_4}
\end{figure}

We study the relation between the rheology and the anisotropic microstructure by measuring the stress responses along the $X$-axis, $\sigma_{xy}$, and the $Z$-axis, $\sigma_{zy}$, simultaneously for different phase angles $\delta$.  We plot $\sigma_{xy}$ and $\sigma_{zy}$ versus $\delta$  in Fig.~\ref{fig:fig_4} (a). Despite the dramatic change in microstructure, we find no measureable difference between $\sigma_{xy}$ and $\sigma_{zy}$ and that both are independent of $\delta$. 

To measure the Pe-dependence of $\sigma_{xy}$ and $\sigma_{zy}$ we perform an amplitude sweep over the range $5\times 10^{-4} \leq \gamma \leq 3.00$ while keeping the frequencies fixed at 31.4 $s^{-1}$. We plot $\sigma_{xy}$ and $\sigma_{zx}$ versus P$\acute{\rm{e}}$clet number 
for both linear and circular polariztions in Fig.~\ref{fig:fig_4}(b). We find quantitatively similar dependencies for both polarizations. These data indicate that the shear induced in-plane structure does not alter the suspension rheology over this range of Pe.

\subsection{Superposition spectroscopy}

\begin{figure}
\includegraphics[width=0.5\textwidth]{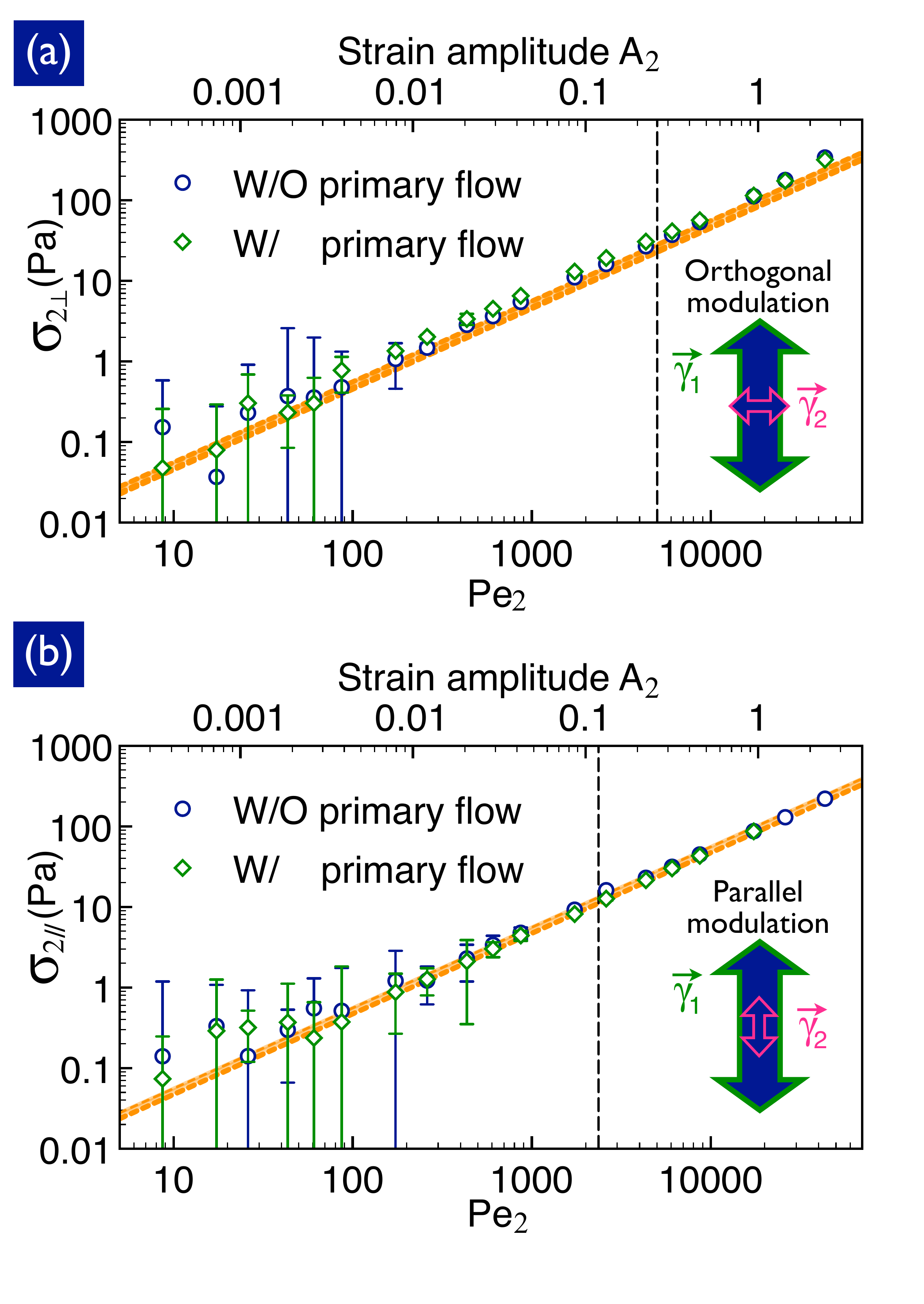}
\caption{Stress response $\sigma_{2 \bot}$ and $\sigma_{2 \parallel}$ measured using superposition spectroscopy for orthogonal (a) and parallel modulations (b). (a) $\sigma_{2 \bot}$ measured in the orthogonal modulation experiment is plotted as a function of \rm{Pe$_2$}. The stress response without the primary flow is also shown in the plot for comparison. The dashed vertical line (black) is the P$\acute{\rm{e}}$clet number of the primary flow, Pe$_1$. The oblique dashed line is the theoretical calculation of the hydrodynamic stress response\cite{Cheng2002}. The narrow orange band represents the uncertainty in the viscosity calculation due to 5$\%$ change in the suspension volume fraction. (b) $\sigma_{2 \parallel}$ measured in the parallel modulation experiment is plotted versus \rm{Pe$_2$}. In the parallel modulation experiment, Pe$_1$ (dashed line) is smaller than that of orthogonal modulation due to the limitation of the piezo travel distance.}
\label{fig:fig_5}
\end{figure}

To probe the anisotropic rheological properties of the suspensions, we perform a superposition spectroscopy measurement \cite{Vermant,orthogonal}. This method has recently gained traction for measuring the shear thinning behavior of polymer melts\cite{Vermant, orthogonal} and colloidal glasses \cite{Ovarlez,Farage}, as well as slow relaxations in granular systems\cite{Ling-Nan}. To perform these measurements we apply a supperposition of shear flows as described in Eq.~\ref{eq:2D} with  frequencies $\omega_1=31.4$$s^{-1}$ for the primary flow and $\omega_2=251$$s^{-1}$ for the secondary flow. Simulatneously we measure the stress response along the $X$-axis and the $Z$-axis using our bi-axial force measurement device.

We conduct experiments using parallel and orthogonal modulations of the primary shear flow. For the orthogonal modulations, we set $\hat{\gamma}_1\cdot \hat{\gamma}_2 = 0$, $\gamma_1=2.50$ and $\gamma_2$ is varied. The primary flow is held fixed with Pe$_1 = 4.73 \times 10^3$ and is used to generate the string phase. The secondary flow is varied over the range $8.60< \rm{Pe}_2<4.30 \times 10^4$ to probe the suspension's rheological response. For the parallel modulation, we set $\hat \gamma_1 \cdot \hat \gamma_2 = 1$, $\gamma_1=1.25$, and $\gamma_2$ is varied. The primary flow is held fixed with Pe$_1 = 2.36 \times10^3$ and is used to generate the string phase. The secondary flow, which is now parallel to the primary flow, is varied over the range $8.60< \rm{Pe}_2<1.95 \times 10^4$ to probe the suspensions rheological response along the flow direction. In both cases we record the total stress for 500 cycles and then take a Fourier transform of the measurement to read out the response - $\sigma_{zy}$, or $\sigma_{xy}$ at the frequency $\omega_2$.

We plot $\sigma_{2 \bot}=\sigma_{zy}(\omega_2)$ for the orthogonal modulation (green diamonds) and $\sigma_{2 \parallel}=\sigma_{xy}(\omega_2)$ for the parallel modulation (green diamonds) versus Pe in Fig.~\ref{fig:fig_5}(a) and (b) respectively. We find that the measurements for both modulations are quantiatively similar. In both cases we find a linear dependence of stress on Pe indicating a Newtonian response. These results suggest that there may not be any effect of the structure on the suspension shear rheology. To test this hypothesis we conduct additional experiments in which the primary flow is absent (blue circles). We find that the stress measurements for these flows are quantitatively similar to those where the primary flow is applied. Collectively these results demonstrate that despite the formation of string structures the suspension response remains isotropic. In addition, we calculate the high frequency limit stress response for an isotropic suspension with $\phi = 0.39$  using \cite{Cheng2002}:
\begin{eqnarray}
\eta_H=\frac{ 1+\frac{3}{2}\phi [1+\phi (1+\phi-2.3 \phi^2)]}{1-\phi[1+\phi(1+\phi-2.3\phi^2)]}. 
\label{eq:viscosity}
\end{eqnarray}

We find that the predicted viscosity $\eta_H=240$mPa$\cdot$s gives a stress response (orange dashed lines in Fig.~\ref{fig:fig_5})that is in excellent agreement with the data. These results imply that for the range of Pe explored, the shear stress response is dominated by a hydrodynamic contribution that is independent of the shear induced suspension structure.  

\section{Conclusion}

Using a bi-axial confocal rheoscope, we measured the rheological response of the vorticity-aligned string phase along the flow and vorticity directions. This apparatus enabled us to control the orientation and morphology of the sheared colloidal suspension under confinement. We showed that by varying the phase between the two shear directions the sample transitions from a string phase to an isotropic phase. By employing various bi-axial shear protocols, we found that despite it's anisotropic structure, the string phase rheology is quantitatively similar to the that of the isotropic suspension. The result clearly shows that hydrodynamic interactions play a crucial role in the formation of the string phase and dominate its rheological response \cite{string}. In addition, they demonstrate that for this volume fraction and degree of confinement, the hydrodynamic contribution to the shear stress is insensitive to the suspension microstructure. Whether such anisotropic structures lead to a normal stress difference remains unknown.

\end{document}